\documentclass[]{myaa}
\usepackage{graphicx}
\usepackage{txfonts}
\usepackage{natbib}
\usepackage{url}
\usepackage{epstopdf} 
\usepackage[figuresright]{rotating}

\usepackage[a4paper,breaklinks]{hyperref}

\def\teff{$T\rm_{eff}$ }
\def\kms {\,$\mathrm{km\, s^{-1}}$ }

\newcommand{\g}{\ensuremath{g}}

\newcommand{\glog}{\ensuremath{\log\g}}
\newcommand{\pun}[1]{\,#1}
\newcommand{\cobold}{\ensuremath{\mathrm{CO}^5\mathrm{BOLD}}}

\newcommand{\beq}{\begin{equation}}
\newcommand{\eeq}{\end{equation}}

\idline{1}{1}
\begin{document}

\title{Sulphur  in the Globular Clusters \object{47 Tucanae} and \object{NGC 6752}
\thanks{Based on
observations made with the 
ESO VLT-Kueyen telescope at the Paranal Observatory, Chile,
in the course
of the ESO-Large Programme 165.L-0263}
}
\subtitle{}

\author{
L. Sbordone  \inst{1,2}\and
M. Limongi   \inst{3}\and
A. Chieffi   \inst{4}\and
E. Caffau    \inst{2}\and
H.-G. Ludwig \inst{1,2}\and
P. Bonifacio \inst{1,2,5}
}
\institute{
CIFIST Marie Curie Excellence Team
\and
GEPI, Observatoire de Paris, CNRS, Universit\'e Paris Diderot; Place
Jules Janssen 92190
Meudon, France
\and
INAF -- Osservatorio Astronomico di Roma, Italy
\and
INAF -- IASF, Rome, Italy 
\and
INAF -- Osservatorio Astronomico di
Trieste, Italy
}
\authorrunning{Sbordone et al.}
\titlerunning{Sulphur abundance in 47 Tuc and NGC 6752}
\offprints{luca.sbordone@obspm.fr}
\date{Received ...; Accepted ...}

\abstract
{ The light elements Li, O, Na, Al, and Mg are known to show
star to star variations in the globular clusters
\object{47 Tuc} and \object{NGC~6752}. Such variations are interpreted
as due to processing in a previous generation of stars.
}
{In this paper we investigate the abundances of the
$\alpha$-element sulphur, for which no previous measurements
exist. In fact this element has not been investigated in any
Galactic globular cluster  so far. The only  globular cluster
for which such measurements are available is \object{Terzan 7},
which belongs to the \object{Sgr dSph}.
}
{We use high resolution spectra of the \ion{S}{i} Mult. 1,
acquired with the UVES spectrograph at the 8.2m VLT-Kueyen 
telescope, for turn-off and giant stars in the
two globular clusters. The spectra are analysed making
use of ATLAS static plane parallel model atmospheres and 
SYNTHE spectrum
synthesis. We also compute 3D corrections from
\cobold\ hydrodynamic models and apply corrections
 due to NLTE effects taken from the literature. 
}
{In the cluster NGC 6752 sulphur has been measured
only in four subgiant stars. We find no significant star to
star scatter and a mean $\rm \left\langle[S/Fe]\right\rangle = +0.49 \pm 0.15$, consistent
with what observed in field stars of the same metallicity.
In the cluster 47 Tuc we measured S in 4 turn-off and 5 subgiant
stars with a mean  $\rm \left\langle[S/Fe]\right\rangle = +0.18 \pm 0.14$.
While this result is compatible with no star to star scatter
we notice a statistically 
significant correlation of {{ the}} sulphur abundance with {{ the}}
sodium abundance and a tentative correlation with {{ the}} silicon
abundance. 
}
{The sulphur -- sodium correlation is not easily
explained in terms of nucleosynthesis. An origin
due to atomic diffusion can be easily dismissed.
The correlation cannot be easily dismissed either,
in view of its statistical significance, until better
data for a larger number of stars is available.}
\keywords{Stars: abundances -- Galaxy: globular clusters: individual: NGC 104 -- Galaxy: globular clusters: individual: NGC 6752}
\maketitle


\section{Introduction}

There is substantial consensus that the so-called $\alpha$-elements (O, Ne, Mg, Si, S, Ar, Ca, and Ti) are synthesized essentially during hydrostatic and explosive burning phases
in massive stars, and then released in the interstellar medium when such stars undergo type II Supernova (SN) explosion { \citep{chieffi04}}. As such, $\alpha$-elements provide crucial { information} to reconstruct the star formation history of stellar populations. O, Mg, Si, Ca, and Ti are easily measured in stellar spectra and are thus the most frequent choice in stellar abundance studies. S on the other hand is the natural choice when abundances are determined in the Interstellar Medium (ISM), Planetary Nebulae (PN), or extragalactic absorption systems (e.g. Damped Lyman Alpha systems, DLA) due to the ease of measuring it and its weak tendency to form dust. The extent to which stellar and ISM / PN /  DLA $\alpha$-element enrichments are comparable is thus dependent on wether or not S is supposed to vary in lockstep with the other $\alpha$-elements. 

Sulphur abundances in Milky Way (MW) stars have been obtained by several authors in the latest years \citep[][among the most recent]{nissen07,caffau2007,ryde06,zolfo05,korn05,nissen04,ryde04}, but agreement on its behavior at low metallicities is not reached yet. While some of the cited authors \citep[e.g.][]{nissen07} find S to follow the same trend as other $\alpha$-elements ([S/Fe] showing a plateau at low metallicities), others point towards a steady growth of [S/Fe] with decreasing metallicity \citep[][]{israelian01,takada02}, or an increase in the { scatter} of [S/Fe] at low [Fe/H] \citep[][]{zolfo05}. The cited debate can in fact be extended to the overall behavior of $\alpha$-elements: while the standard view of [$\alpha$/Fe] evolution with metallicity has for some time now assumed the presence of a plateau below { [Fe/H]$\sim$--1} \citep[][]{mcwilliam97}, observations as well as models suggest the possibility of a continuing rise down to the lowest observed metallicities, at least for Mg  and O (first pointed out by \citealt{abia89}, and more recently by \citealt{francois04}, and \citealt{jonay08}).

It is perhaps surprising that sulphur has been studied up to now almost { exclusively} in field stars. To our knowledge, the only analysis of { the} S abundance in globular clusters is also the only one for extragalactic stars \citep[][in \object{Terzan 7}, one of the globular clusters of the Sagittarius dwarf Spheroidal Galaxy]{STer7}, where it behaves consistently with the rest of the $\alpha$-elements studied in the cluster, showing a low ratio with iron typical of dwarf spheroidal systems \citep[][]{tarantella04,sbordone07}. But on the other hand, S measurements are totally missing in MW Globular Clusters (GC). Globular Clusters are somewhat peculiar environments regarding chemical enrichment, displaying some peculiar abundance anomalies and abundance correlations which have never been found elsewhere \citep[for a review see][]{gratton04}. Among them the most well known is likely the so-called O-Na anticorrelation, which appears to be widespread in MW globular cluster stars, and the Mg-Al anticorrelation. The nucleosynthetic processes  responsible for this are believed to be essentially understood, since the chemical yields match well the ones derived from p-capture processes occurring in (or near) regions where H is being burned via { the} CNO cycle at sufficiently high temperatures \citep[][]{gratton04}. On the other hand, such a condition is reached during different phases of stellar evolution, so that the actual processing site is still debated, as well as the mechanism  through which the enriched gas is dispersed inside the cluster \citep[][]{ventura02,dantona05,decressin}. There is nevertheless some consensus that the yields should be released through slow moving ejecta, thus allowing the GC { gravitational} potential to retain them more easily. 

We present here the first sulphur abundances in the GC \object{47 Tuc} and \object{NGC 6752}, both of which display evident Na-O and Mg-Al abundance anticorrelations. They are also the only two clusters where { a} Na-Li anticorrelation has also been observed \citep[][]{pasquini05,B07}.


\section{Observational Data }

\begin{figure}
\begin{center}
\resizebox{\hsize}{!}{\includegraphics[]{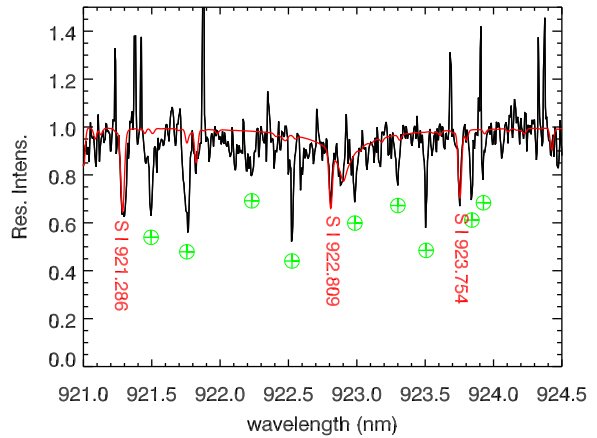}}
\caption{\label{spsam} The (normalized) spectrum of \object{47 Tuc} star $\#$ 1012 with superimposed a synthetic spectrum with \teff~ 5832 K, \glog~ 4.05, [Fe/H]=-0.5. The three lines of \ion{S}{i} mult. 1 are labeled, as well as the most prominent telluric absorptions (marked with $\oplus$ symbol).}
\end{center}
\end{figure}

The spectra analysed in this paper have been  acquired in the course  of the ESO-Large Programme 165.L-0263. Observations, atmospheric 
parameters and abundances for elements other than S have been presented in \citet{gratton01}, \citet{carretta04}, and \citet{B07}. We are here concerned only with the dichroic \#2, CD4 cross-disperser UVES@VLT \citep[][]{dekker00} spectra (setting centered at 760 nm). For most observations the slit width was of $1''$ for a resolution of R$\sim$43000. For the purpose of { the} sulphur measurement, only upper-red CCD data were used. Typical Signal-to-Noise ratios (S/N) in the range were of about 15-20 { per resolution element}. { Sulphur abundances were determined for nine stars observed in \object{47 Tuc}, four of them being turn off { (TO)} stars, 5 { subgiants (SG)}. Four stars, all { subgiants}, were considered in \object{NGC 6752}}. { Typical exposure times were 1h per star for \object{NGC 6752} SG stars, 2h per stars for \object{47 Tuc} SG and 4h per stars (in both cases performing multiple 1h exposures) for \object{47 Tuc} TO stars}. { An example of \ion{S}{i} mult. 1 region is shown in fig. \ref{spsam}.}
  

\section{Analysis}

The atmospheric parameters have been taken from \citet{gratton01} for the stars in \object{NGC 6752} and from \citet{carretta04}  and \citet{B07} { (for star $\#$ 952)} for the stars in \object{47 Tuc}. { \object{47 Tuc} stars have a typical temperature of  5800 K for the TO stars ($\log$ g=4.05) and of 5100 K for the subgiants ($\log$ g=3.84), the four \object{NGC 6752} share the same atmosphere parameters, \teff=5347 K $\log$ g=3.54.} For each star we computed a model atmosphere using  version 9 of the ATLAS code  \citep{kurucz,k05} running under Linux  \citep{sbordone,s04}. We used the updated Opacity  Distribution Functions of \citet{c_k} with microturbulent velocity of 1 \kms and enhancement of $\alpha$-elements. The synthetic spectra were computed using the SYNTHE suite \citep{kurucz,k05} running under Linux \citep{sbordone,s04}.

The strongest lines of \ion{S}{i} in the optical range are the lines of Mult. 1 (three lines around 923 nm) and of Mult.6 (two lines at 869 nm, see Table \ref{byline}). We were unable to convincingly detect the lines  of Mult. 6 in any of our spectra, we therefore did not even search for Mult. 8 ( around 675 nm) which is weaker.  Our analysis is therefore based exclusively on the lines of Mult. 1. We adopt the   $\log gf$ values of the  NIST database, which  holds the values of \citet{wiese}, which are experimental but of D quality, corresponding to a possible error up to 50 \%  (see Table \ref{byline} for the values). A thorough discussion of the available values for  \ion{S}{i} atomic data is avilable \citet{zolfo05}.

The lines are in a spectral range contaminated by telluric absorption, so that they are rarely all free from telluric contamination. Moreover, due to the low metallicity of the two clusters examined and to the low S/N ratio of the spectra, in most cases, only one of the \ion{S}{i}  lines of Mult. 1 is detectable. Due to the radial velocity of the clusters, the line which happens to be the most often unblended is the one at 922.8\pun{nm}, that is the line laying on the wing of the hydrogen Paschen $\zeta$ line. The  measurement of the equivalent width (EW)  of lines falling on the wings of broad features is not easy, and produces a value which is usually lower, with  respect to the true EW, by about 5\% \citep{zolfo05}. To determine the sulphur abundance a line profile fitting procedure, taking into account the profile of the Paschen $\zeta$   is to be preferred. The other two lines of Mult. 1 are far enough from the Paschen $\zeta$ not to be affected by it, so that for these lines the EW measurement would be a good choice. To keep the analysis as homogeneous as possible we nevertheless decided to determine the sulphur abundance using line profile fitting for all the unblended features. Such fitting was performed by means of the code described in \citet{zolfo05}. { In the fitting procedure, the line broadening was kept fixed at the instrumental level (7 km/s). At this resolution, and with this S/N ratio, typical macroturbulent and/or rotational broadening would not affect the measurement significantly.}
We discarded the lines contaminated by telluric absorption because the telluric line subtraction from a spectrum decreases the S/N that is already low. { Also, we did not possess fast rotator spectra taken at the same time as the science exposures, which would have imposed to rescale the telluric features, an operation which would add further uncertainty to the measurement.} Line-by-line S abundances are listed in Table \ref{byline}.

\begin{table}
\caption{Sulphur line-by-line abundances for the program stars. The wavelengths and log gf if the three lines are indicated as well.\label{byline}
}
\centering
\begin{tabular}{rcrrrr}
\hline
 $\lambda$ &  921.286 nm & 922.809 nm &  923.754 nm \\
 $\log$ gf & 0.42 & 0.26 & 0.04 \\
\hline\hline
\noalign{\smallskip}
\object{NGC 6752} & & & \\

\hline
  \object{1406} &  5.80 &  6.07  &  6.04  \\
  \object{1665} &  --    &  6.35  &  6.29  \\
  \object{1461} &  --    &  6.35  &  6.37  \\
  \object{1481} &  --    &  6.35  &  6.36  \\
\hline
\noalign{\smallskip}
\object{47 Tuc} & & & \\

\hline
  \object{1012} &  --    &  6.56  &  --    \\
  \object{1081} &  --    &  6.68  &  --    \\ 
   \object{952} &  --    &  6.63  &  --    \\
   \object{975} &  --    &  6.72  &  --    \\
\object{201075} &  6.65 &  6.86  &  6.43 \\
\object{206415} &  6.73 &  --     &  --    \\
   \object{429} &  --    &  6.87  &  --    \\
   \object{433} &  --    &  6.87  &  --    \\
   \object{482} &  --    &  6.80  &  --    \\
\hline
\end{tabular}
\end{table}

The \ion{S}{i} lines of Mult. 1 are known to suffer from  departures from LTE \citep{takeda}. We therefore  applied the NLTE corrections, interpolated in  table 2 of \citet{takeda} to take this effect into account. For our programme stars these are all negative and range between --0.1 and --0.3 dex.

{ \subsection{Estimation of analysis uncertainties}
\label{uncert} 

Uncertainty in the S abundance determination stems from both the limited S/N ratio of the spectra and the systematics of the analysis. To determine the effect of S/N we followed the same method described in \citet{zolfo05}, namely by  performing Monte Carlo simulations. In that paper the error estimates are obtained for a sample of dwarf stars with [Fe/H]=--1.5. We have performed additional Monte Carlo tests to assess the \teff sensitivity of the S lines in the SG stars (\teff being the only parameter showing a significant difference between TO and SG stars), and to ascertain that such uncertainties apply to the more metal rich objects of \object{47 Tuc} and came up with values which are essentially identical to the ones in \citet{zolfo05}, which we will then assume from now on.

Concerning random errors, the result of a 1000-events simulations, for a parameter set compatible with the TO stars set, indicates an 1$\sigma$ uncertainty (for a single line) of 0.12 dex for S/N=30, and 0.22 dex for S/N=15.  The same values hold for the SG stars. Given the typical S/N of our spectra, we will assume henceforth a typical uncertainty of 0.2 dex to be due to the S/N.

For the TO sample, a change in \glog~ of $\pm$0.25 dex causes a $\pm$ 0.08 dex variation in A(S), $\pm$ 100 K in temperature cause a variation of $\mp$0.06 dex in A(S), while [Fe/H]$\pm$0.2 dex induces a $\pm$ 0.03 dex variation in A(S). Comparing with table 3  of \citet{carretta04} one notices that both \teff and \glog~ variations act in opposite senses on [Fe/H] and on A(S). On the one hand, \glog~ has a maximum stated error in \citet{carretta04} of 0.08 dex. Taking into account both the effect on metallicity and the one on A(S), this would lead to a variation of 0.03 dex in [S/Fe], negligible in the present scope. 

On the other hand, \citet{carretta04} claim a maximum \teff error of $\pm$ 90 K for TO stars and $\pm$60 K for subgiants. Shifting the whole temperature scale rigidly would not make or break the [S/Fe] spread we observe, but one could argue that the \citet{carretta04} \teff scale might be excessively ``compressed''. By {\em lowering} the TO stars temperature and {\em rising} the one of the subgiants by the maximum amount stated, one would reduce significantly the observed [S/Fe] spread. A 90 K lower  temperature for the TO stars would lower their [Fe/H] by 0.08 dex. The effect on S would then be due both to the reduced \teff and to the decreased metallicity, the total effect being roughly $\Delta$A(S)=+0.04 dex. A similar calculation for the SG stars leads to $\Delta$A(S)=--0.03, which become $\Delta$[S/Fe]=+0.12 for the TO and --0.09 for the SG. This would reduce the [S/Fe] spread by 0.21 dex, eliminating something less than half of it.
On the other hand, such a change would lead [Fe/H] to differ between TO and SG stars. Currently, the two subgroups give precisely the same metallicity (--0.67) with a dispersion of 0.02 dex among TO stars and 0.01 dex among SG stars. Applying the aforementioned opposite temperature differences would separate the TO and SG stars by 0.14 dex in metallicity, roughly five times the quadratic sum of the two group uncertainties. We therefore suggest that any \teff bias is very likely to be a simple additive shift  to both TO and SG stars temperature. It is also worth noting that some imbalance ($\sim$0.13 dex) exists between \ion{Fe}{i} and \ion{Fe}{ii} abundances in the giants sample of \citet{carretta04}. According to Carretta and collaborators' error budget, this would be entirely solved by rigidly increasing the \teff in subgiants by 100 K. Again, a rigid shift of the \teff scale would not affect the spread in A(S) nor in [S/Fe], and would also break the ionization equilibrium in the TO sample. Shifting the giant sample only would reduce the spread in sulphur abundance, but generate a Fe abundance spread. Besides, any increase in the subgiants \teff would mainly affect \ion{Fe}{i}, raising its abundance. A discrepancy already exists between \citet{carretta04} and \citet{koch08} [Fe/H] scales, \citet{koch08} deriving an average [Fe/H] {\em lower} by 0.09 dex. Such a discrepancy would be worsened, should the temperature for the \citet{carretta04} subgiants be increased.

}

\subsection{3D corrections}
We computed 3D corrections for the \ion{S}{i} 922.8 nm line, for a set of atmosphere models encompassing the parameters of our program stars. In Table \ref{3dcor} we present the resulting values. Assuming an equivalent width of 12 pm (typical for the observed stars), we tabulate the difference between the abundance needed  to reproduce the { Equivalent Width (EW)} in the 3D CO$^5$BOLD model \citep[][]{frey02,wede04} and the one needed when using the comparison 1D$_{LHD}$ model \citep[][]{caffau07P}. As such, the 3D abundance is obtained by {\em adding} the correction to our 1D abundances. Table \ref{3dcor} reports the model parameters as well. 3D corrections appear to be on average 0.15 dex and positive, and appear highly homogeneous across the explored parameter space. { In particular, the differences in the corrections for TO and SG stars are small.} They have {\em not} been added to our measurements in the table and in the figures, since comparison elements, in this and other studies do not have 3D corrections applied.

\begin{table}
\caption{3D corrections for the \ion{S}{i} 922.8 nm line, computed for a set of stellar parameters compatible with the ones in our sample, for an EW of 12 pm. The corrections are in the form Abu(3D)-Abu(1D). The model parameters are listed as well. A microturbulence of 1.5 km s$^{-1}$ has been employed in the LHD 1D model. \label{3dcor}}

\centering
\begin{tabular}{lrrr}
\hline
\teff & $\log g$ & [Fe/H] & Corr. \\
\hline\hline
5000 & 4.0 & 0.0 & 0.141 \\
5000 & 4.0 & --1.0 & 0.140 \\
4920 & 3.5 & 0.0 & 0.146 \\
4930 & 3.5 & --1.0 & 0.167 \\
\hline
\end{tabular}
\end{table}


\section{Results}

\begin{table}
\caption{Sulphur abundances for the subgiant stars in \object{NGC 6752}. The star identification comes from \protect\citet{gratton01}, as well as the atmosphere parameters: \teff=5347 K $\log$ g=3.54 V$_{\mathrm{turb}}$= 1.1 \kms , [Fe/H]=-1.43 for all stars. The abundances of Na and Mg are from \protect\citet{gratton01}. The number of \ion{S}{i} lines used is given in the second column.\label{ngc6752}
}
\centering
\begin{tabular}{rcrrr}
\hline
Star & N & [S/Fe] & [Na/Fe] & [Mg/Fe] \\
\hline\hline
 \object{1406}&3 &+0.19 & +0.02 & $+0.20$ \\
 \object{1665}&2 &+0.54 & +0.10 & $+0.10$ \\
 \object{1461}&2 &+0.58 & +0.29 & $+0.10$ \\
 \object{1481}&2 &+0.57 & +0.54 & $-0.07$ \\
\hline
\end{tabular}
\end{table}

\begin{table*}
\caption{Sulphur abundances for the stars in \object{47 Tuc}. The star numbers come from \protect\citet{carretta04}, as well well as atmospheric parameters and abundances of Na, Mg and Si. For star \# 952 see \protect\citet{B07}. The number of \ion{S}{i} lines used  is given in the fourth column.\label{47tuc}}
\centering
\begin{tabular}{rrrrrrrrr}
\hline
 Star     &\teff  & log g & N  & [S/Fe]   & [Fe/H]  & [Na/Fe] & [Mg/Fe] &[Si/Fe] \\   
          &  K    &  cgs  &    &   dex    &   dex   &   dex   &   dex   & dex \\
\hline\hline
     \object{1012}  & 5832  & 4.05  & 1  & $ -0.01$ & $-0.64$  & $-0.14$  & $+0.45$ &$ +0.03$ \\
     \object{1081}  & 5832  & 4.05  & 1  & $ +0.11$ & $-0.64$ & $-0.34$  & $+0.50$ &$ +0.13$ \\
        \object{952} & 5832  & 4.05  & 1  & $ +0.06$ & $-0.64$ & $-0.08$  & $         $ &$         $ \\
       \object{975}  & 5832  & 4.05  & 1  & $ +0.15$ & $-0.64$ & $+0.22$ & $+0.21$ &$ -0.10$ \\
 \object{201075}  & 5165  & 3.84  & 3  & $ +0.12$ & $-0.68$ & $+0.11$ & $+0.42$ &$ +0.38$ \\
 \object{206415}  & 5112  & 3.84  & 1  & $ +0.22$ & $-0.70$ & $+0.10$ & $+0.34$ &$ +0.45$ \\
       \object{429}  & 5081  & 3.84  & 1  & $ +0.31$ & $-0.65$ & $+0.31$ & $+0.39$ &$ +0.42$ \\
       \object{433}  & 5106  & 3.84  & 1  & $ +0.44$ & $-0.78$ & $+0.24$ & $+0.49$ &$ +0.21$ \\
       \object{482}  & 5090  & 3.84  & 1  & $ +0.21$ & $-0.62$ & $+0.06$ & $+0.64$ &$ +0.28$\\
\hline
\end{tabular}
\end{table*}

\begin{figure}
\begin{center}
\resizebox{\hsize}{!}{\includegraphics[]{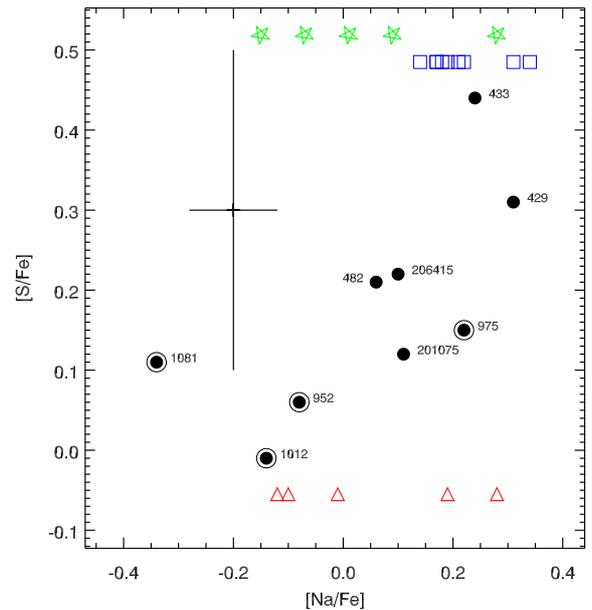}}
\caption{\label{nas} [S/Fe] plotted against [Na/Fe] for our sample of \object{47 Tuc} stars. Na spread is added for reference from \protect\citet{brown92} (open red triangles), \protect\citet{alvesbrito05} (open green stars), and \protect\citet{koch08} (open blue squares). A typical error for our dataset is also displayed. Stars in our sample belonging to the TO are identified by a larger open circle around them.}
\end{center}
\end{figure}

Our results are summarised in Tables \ref{ngc6752} and \ref{47tuc}. For the reader's information we report there also some information on our program stars extracted from \citet{gratton01} and \citet{carretta04}. { The
other abundances coming from these papers are always 1D-LTE values, with the exception of Na, which had NLTE corrections applied according to \citet{gratton99}.} { We adopted a solar sulphur abundance of A(S)=7.21 in the [S/Fe] determination.}

In NGC 6752, due to its low metallicity { ([Fe/H]=$-$1.43)},  we were only able to measure the sulphur abundance in the four subgiant stars. In the fainter TO stars the S/N ratio was too low to detect the \ion{S}{i} lines. From the four subgiants we derive $\rm <[S/Fe]> = +0.49 \pm 0.15$. Our estimated error on the sulphur abundances is 0.2 dex,  dominated by the noise in the observed spectra.  Thus the dispersion around the mean is fully compatible with the errors. Moreover, [S/Fe] in \object{NGC 6752} is fully compatible with the value found among field stars at this metallicity (see \citealt{zolfo05} figure 10).

\begin{figure}
\begin{center}
\resizebox{\hsize}{!}{\includegraphics[]{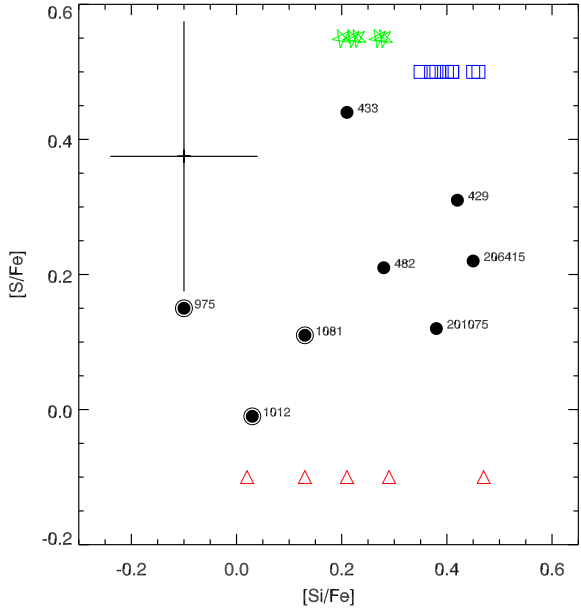}}
\caption{\label{sis} [S/Fe] plotted against [Si/Fe] for our sample of \object{47 Tuc} stars.  Si spread is added for reference from \protect\citet{brown92} (open red triangles), \protect\citet{alvesbrito05} (open green stars), and \protect\citet{koch08} (open blue squares). A typical error for our dataset is also displayed. Stars in our sample belonging to the TO are identified by a larger open circle around them.}
\end{center}
\end{figure}

For the more metal-rich cluster, \object{47 Tuc},  we have been able to measure at least one \ion{S}{i} line for all the stars. In star \object{201075} all the three lines of Mult. 1 were measurable. The [S/Fe] ratio varies by almost 0.5 dex, with an average value of $\rm <[S/Fe]>=0.18\pm 0.14$.  Given that the uncertainty in { the} S abundance in each star is estimated at 0.2 dex, the observed spread can in fact simply reflect the measurement uncertainty. Moreover, as in the case of \object{NGC 6752}, the mean [S/Fe] is totally compatible with the value displayed by field stars of abundance comparable to \object{47 Tuc} \citep{zolfo05}. 

\object{47 Tuc} is believed to display Na-O as well as Mg-Al abundance anti-correlations { \citep{carretta04}}. We thus decided to check for possible correlations between S and some of the cited elements. Sulphur has never been measured before in \object{47 Tuc} but interesting insight can be gained regarding other elements by comparing \citet{carretta04} 
with the ones obtained by \citet{brown92}, \citet{alvesbrito05}, and \citet{koch08}. In Fig. \ref{nas} [S/Fe] is plotted against [Na/Fe]. Black filled circles represent our data, points surrounded by a slightly larger circle indicate TO stars, and every data point is flagged with the corresponding star number. The typical size for the error is also indicated. \citet{brown92}, \citet{alvesbrito05}, and \citet{koch08} do not measure sulphur, but for reference we add their values for [Na/Fe]: blue open squares for \citet{koch08} red open triangles for \citet{brown92}, and green open stars for \citet{alvesbrito05}.
The correlation between Na and S abundance is clearly apparent in our data, and a Kendall $\tau$ rank correlation test in fact determines that the likelihood of [Na/Fe] and [S/Fe] to be correlated is 97.8\%\footnote{{ This is equivalent to say that the likelihood of the null hypothesis is 2.2 \%. This is the likelihood that random errors applied to otherwise uncorrelated distributions might produce a level of correlation ($\tau _{K}$) equal or superior to the observed one.}}. Another interesting thing to notice is how all the studies with the exception of \citet{koch08} detect a significant spread in { the} Na abundance.

In Fig. \ref{sis} we instead plot [S/Fe] versus [Si/Fe], all the symbols being the same as in Fig. \ref{nas}. Again, there is an evident suggestion of a correlation, with the exception of star \object{433} which, while having a very high S abundance, presents a less extreme [Si/Fe]. Likelihood of the correlation from Kendall $\tau$ test is here about 79\%, but grows to 91\% if star \object{433} is removed from the sample. By removing star \object{433} from the [Na/Fe] vs [S/Fe] correlation test, one gets on the other hand a slight {\em decrease} of the correlation likelihood, down to 94\%. It is worth noting how star \object{433} has, according to \citet{carretta04}, an unusually low [Fe/H] (0.14 dex below the average, way beyond the scatter of the \ion{Fe}{i} abundances), and an unsatisfactory \ion{Fe}{i}-\ion{Fe}{ii} ionization equilibrium (0.21 dex difference). Since no evidence ever emerged of \object{47 Tuc} showing a spread in iron abundance, one might regard these results as suggestive of some problem in the analysis of star \object{433}. If we tentatively assign the average Fe abundance ([Fe/H]=--0.66) to star \object{433}, we obtain [S/Fe]=0.32, more in line with the rest of the sample. In fact, star \object{433} has a [S/H] exceeding by just 0.14 dex the average value ($\rm \left\langle[S/H]\right\rangle$=--0.48. [S/H]$_{433}$=--0.34).

{ Given the fact that most low-S stars (and low-Na stars) are TO stars, it is worth investigating whether diffusion in the stellar atmosphere might be responsible for the observed abundance spread and its correlation with the Na abundance. One would in this case expect TO stars to be most affected by diffusion, while SG stars should not display them anymore due to the deeper reach of their atmospheric convective zone. According to current models this is not a viable explanation of the observed spreads. According to \citet{korn07} (fig. 1), by using a model with radiative acceleration and turbulent mixing, an old [Fe/H]=--2 TO star shows photospheric depletion of about 0.25 dex in Na, 0.20 dex in S and 0.15 dex in Fe. The effect would be then 0.05 dex on [S/Fe] and 0.1 dex in [Na/Fe], too small to explain the observations.}


\section{Discussion}

The sulphur abundance in \object{NGC 6752} agrees with the general trend of $\alpha$-elements in the field, and does not present any significant spread. In \object{47 Tuc} the situation is less clear. The presence of a significant correlation between the S and Na abundance, and the hint of a S/Si correlation, suggest that the S abundance spread might indeed be real.

Recently \citet{koch08}, in a thorough analysis of \object{47 Tuc}, based on high quality spectra of bright giants, found only a very limited spread in abundances, and were unable to confirm the presence of a significant spread in O, Na, Al and Mg. { Such discrepancy with the results of}  \citet{brown92}, \citet{carretta04}, \citet{alvesbrito05}, and \citet{B07} { has not found to date a satisfactory explanation.}  { This  is not the venue to reassess in detail the work of the cited authors. Suffice to say that, applying an error analysis similar to the one above sketched for sulphur, the reality of the spreads in \object{47 Tuc} as well as their correlation appear difficult to dismiss. } 

\begin{figure}
\begin{center}
\resizebox{\hsize}{!}{\includegraphics[]{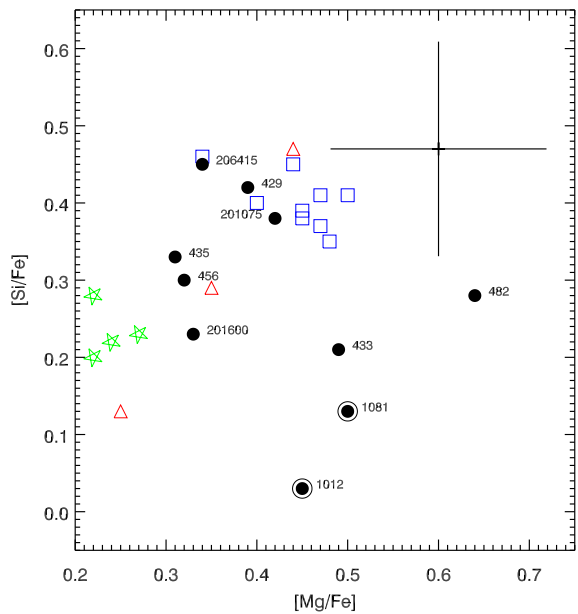}}
\caption{\label{simg} [Si/Fe] plotted against [Mg/Fe] for our sample of \object{47 Tuc} stars, plus \protect\citet{brown92} (open red triangles), \protect\citet{alvesbrito05} (green stars), and \protect\citet{koch08} (open blue squares).  Stars in our sample belonging to the TO are identified by a larger open circle around them. A typical error for our sample is also displayed.}
\end{center}
\end{figure}

In Fig.\ref{simg} we plot [Si/Fe] versus [Mg/Fe] for our stars \citep[abundances are taken from][]{carretta04}, the sample of \citet{brown92} and the one of \citet{koch08}. Symbols are the same as in Fig. \ref{nas} and \ref{sis}. The first thing that can be noted here is that, with the exception of star \object{482}\footnote{ star \object{482} is indeed suspect in a more general way, presenting the largest \ion{Fe}{i} - \ion{Fe}{ii} imbalance of the whole \citet{carretta04} sample.} , our sample and the one of \citet{koch08} present in fact the same spread in [Mg/Fe], but a very different one in [Si/Fe]. { A word of caution should be here given regarding silicon: it appears that Si shows a systematic discrepancy between TO and RGB in extremely metal poor stars \citep{mcwilliam97,cohen04,bonifacio09}. The reasons of the discrepancy are not known, NLTE effects being a possibility. The sense and amount of the discrepancy is compatible with the one seen here, so there { exists} the possibility that the Si spread is not real. On the other hand, such an effect has never been reported on stars as metal rich as the ones in \object{47~Tuc}.} 
Another element for which the various authors do not reach a satisfactory agreement is Ca. Average [Ca/Fe] is 0.34 in \citet{koch08},  0.18 in \citet{carretta04}, but solar according to \citet{brown92} and \citet{alvesbrito05}. 

Another important aspect is of course constituted by the existence of significant correlations among the abundances. 
One might consider { as worrying the fact} that all the TO stars of the sample exhibit a low S abundance. On the other hand, star \object{201075}, which is a subgiant, and the one star where all the three lines of Mult. 1 were measured, shows low S as well. By examining Table \ref{byline} one might notice that the line at 922.8 nm gives the highest abundance for star \object{201075}. Since this line is the one most frequently measured in the other stars, one might object that, had it been the only one measured, the star would have assigned an higher abundance, compatible with the other subgiants. It has although to be noted that in \object{NGC 6752} both this line and the one at 923.7 nm are systematically measured, and no sign exist of the latter giving a lower abundance. { As detailed in sect. \ref{uncert}, the quoted \teff uncertainty would in principle be sufficient to remove the spread between TO and subgiant stars, if TO stars temperature was lowered and SGB stars temperature increased. On the other hand, this would make star \object{201075} even more discrepant, and induce a Fe abundance spread between TO and SGB stars.}  

 { Furthermore, a Monte Carlo simulation of 1000 events has been performed to check the likelihood that a significant correlation might emerge in the [S/Fe] vs. [Na/Fe] plot when the quoted errors are superimposed to a {\em constant} [S/Fe] for all the stars. Unsurprisingly, the result ($\sim$2\% likelihood) confirms the result of the Kendall $\tau$ null-hypothesis probability. It is very unlikely that random errors might have produced the observed Na / S correlation. }

In short,  we believe that strong hints of a { S abundance inhomogeneity} in \object{47 Tuc} do exist. It is nevertheless clear that the currently available data do not allow, due to a combination of low statistics and significant uncertainties, to clarify the matter. Further study would thus be in order, specifically targeting a large number of stars from TO to RGB, and focusing on the elements that show significant spreads.
{ Still, being unable to justify the current observations by means of either random or systematic uncertainties, we feel compelled to look for a nucleosynthetic explanation.} { The} Na-O anticorrelation is usually explained by self pollution within the cluster by slow Na-rich winds, either from AGB stars or from massive rotating stars. In general, the stars of pristine cluster composition are expected to show {\em low} Na abundance, while a higher abundance is indication of  ``polluted'' stars. Should Si and S abundance spreads be confirmed, and should they both correlate with the Na spread { (which is quite uncertain in the Si case)}, this would pose at least two problems. 

First, since Na-poor stars are Si-poor and S-poor as well, the aforementioned model would indicate them as the ``unpolluted'' stars. \object{47 Tuc} would then have formed with an unusual low abundance of Si and S, at variance e.g. with Mg. Low [$\alpha$/Fe] ratios are indeed observed in dwarf galaxies, but typically attributed to environmental effects \citep[e.g.][]{lanfranchi07}, and no evidence exist that they might privilege one explosive product (e.g. Si) over another (e.g. Mg).

Second, as just noted, both Si and S are believed to be, almost exclusively, products of SN II explosive burnings. Elementary models forbid globular clusters to be chemically enriched by internal SN explosions, since the total released energy is comparable to the cluster binding energy, and even one single SN II explosion should remove all the available gas from the cluster potential well, and stop the star formation. This is the reason why ``slow wind'' mechanisms have been invoked to produce e.g. the Na pollution. At the same time, the lack of any detectable spread in Fe abundance in \object{47 Tuc} constitutes an additional problem, since one would expect some amount of Fe to be produced by the SN along with $\alpha$-elements. On the other hand, such simplistic considerations provide likely too crude a model of the actual phenomenon, since they assume a spherically symmetrical, central explosion which energy is efficiently transferred to the cluster gas.  SN II explosion are in fact rarely spherically symmetrical \citep[][]{wang08}. Instead, they tend to release a large part of the energy along more or less collimated jets which are not efficient in transferring the energy to the cluster ISM.  Moreover, SN which explode in a non-spherically symmetrical way tend to produce larger fallback, which would in turn limit their Fe yield, thus eliminating the ``missing Fe spread'' problem. 
It is also suggestive that detailed models of stellar superclusters foresee a destructive interference among shock waves from different supernovae, within a given critical radius \citep[][]{wuensch08}. As a result the inner part of the cluster would not only retain the yields, but also experience an accelerated star formation. Whether this might be the case on the smaller scale of normal GC (or whether some/all the GC might form as superclusters) is anyway uncertain. 
Core collapse SN are a source of Na as well, and we see that { the} Na abundance correlates well with { the} S abundance. The { applicability} of the aforementioned model thus relies on the (yet unexplored) possibility that high Na-yield and high $\alpha$-yield (or at least high S-yield) happen together. 

An alternate way to produce S and Si would be through proton capture reactions. This is the same source invoked to explain { the} Na overabundance, which would be an obvious advantage. On the other hand, S should this way be produced through the $\rm ^{31}P(p,\gamma)^{32}S$ reaction. The phosphorus abundance in \object{47 Tuc} is unknown, and in general, knowledge about P production channels and abundances in stars are sketchy at best \citep[see][and references therein]{caffau07P}. However, P in the Sun is roughly 3 dex less abundant than S. If the same abundance ratio holds at lower metallicities, it would be impossible to produce any significant amount of S at expenses of P. Phosphorus should then be formed at expenses of Si, which should then {\em anticorrelate} with S, which is not the case. The next abundant enough seed is Mg (Mg $\to$ Al $\to$ Si $\to$ P $\to$ S), which indeed might show weak signs of anti-correlation with Si (Fig. \ref{simg}), although they are below any statistical significance.  Given the very scarce knowledge about P abundances and production mechanisms, one might even suppose the ``original'' \object{47 Tuc} P abundance to be heavily enhanced, thus providing enough seeds to form $\sim$0.2 dex of S. {  P abundance has never been measured in \object{47 Tuc}, but its abundance has been determined very recently in Horizontal Branch (HB) stars in \object{NGC 6397} and \object{NGC6752} \citep{hubrig09}, where it appears strongly enhanced, by more than 2 dex in \object{NGC 6752}, and almost 3 dex in \object{NGC 6397}. Chemical anomalies in HB stars are usually attributed to diffusion effects, but the current models \citep{michaud08} do not seem to be account for more than $\sim$0.7 dex of P enhancement in the photospheres of HB stars of similar temperature. We thus think that the possibility of an enhancement of P abundance in globular clusters cannot be ruled out at the moment.}
\\


\begin{acknowledgements}
  { We whish to thank the anonymous referee for the useful comments who really helped to improve the paper.}
  The authors L.S., H.-G.L., P.B. acknowledge financial
  support from EU contract MEXT-CT-2004-014265 (CIFIST). { This research has made use of the SIMBAD database, operated at CDS, Strasbourg, France, and of NASA's Astrophysics Data System.}
\end{acknowledgements}

\bibliographystyle{aa}

\begin{thebibliography}{}

\bibitem[Abia 
\& Rebolo(1989)]{abia89} Abia, C., \& Rebolo, R.\ 1989, \apj, 347, 186 


\bibitem[Alves-Brito et 
al.(2005)]{alvesbrito05} Alves-Brito, A., et al.\ 2005, \aap, 435, 657 

\bibitem[Bonifacio et al.(2009)]{bonifacio09} Bonifacio, P., et 
al.\ 2009, arXiv:0903.4174 


\bibitem[Bonifacio et al.(2007)]{B07} Bonifacio, P., et al.\ 2007, \aap, 470, 153 

\bibitem[Brown 
\& Wallerstein(1992)]{brown92} Brown, J.~A., \& Wallerstein, G.\ 1992, \aj, 104, 1818 

\bibitem[Caffau et al.(2007)]{caffau2007} Caffau, E., Faraggiana, 
R., Bonifacio, P., Ludwig, H.-G., \& Steffen, M.\ 2007,  
A\&A in press, arXiv:0704.2335 

\bibitem[Caffau \& Ludwig(2007)]{zolfito} Caffau, E., \& 
Ludwig, H.-G.\ 2007, \aap, 467, L11

\bibitem[Caffau et 
al.(2007)]{caffau07P} Caffau, E., Steffen, M., Sbordone, L., Ludwig, H.-G., \& Bonifacio, P.\ 2007, \aap, 473, L9

\bibitem[Caffau et al.(2005a)]{STer7} 
Caffau, E., Bonifacio, P., Faraggiana, R., \& Sbordone, L.\ 2005, \aap, 436, L9 

\bibitem[Caffau et al.(2005b)]{zolfo05} Caffau, E., Bonifacio, 
P., Faraggiana, R., Fran{\c c}ois, P., Gratton, R.~G., \& Barbieri, M.\ 
2005, \aap, 441, 533 

\bibitem[Carretta et al.(2004)]{carretta04} 
Carretta, E., Gratton, R.~G., 
Bragaglia, A., Bonifacio, P., \& Pasquini, L.\ 2004, \aap, 416, 925 

\bibitem[{{Castelli} \& {Kurucz}(2003)}]{c_k}
{Castelli}, F. \& {Kurucz}, R.~L. 2003,
in IAU Symposium 210, 
``Modelling of Stellar Atmospheres'' ed. N.~{Piskunov},
  W.~W. {Weiss}, \& D.~F. {Gray}, 20P
  
 \bibitem[Chieffi 
\& Limongi(2004)]{chieffi04} Chieffi, A., \& Limongi, M.\ 2004, \apj, 608, 405 

\bibitem[Cohen et al.(2004)]{cohen04} Cohen, J.~G., et al.\ 
2004, \apj, 612, 1107 

\bibitem[D'Antona et al.(2005)]{dantona05} D'Antona, F., 
Bellazzini, M., Caloi, V., Pecci, F.~F., Galleti, S., 
\& Rood, R.~T.\ 2005, \apj, 631, 868 

\bibitem[Decressin et al.(2007)]{decressin} 
Decressin, T., Charbonnel, C., \& Meynet, G.\ 2007, \aap, 475, 859 

\bibitem[Dekker et al.(2000)]{dekker00} Dekker, H., D'Odorico, 
S., Kaufer, A., Delabre, B., \& Kotzlowski, H.\ 2000, \procspie, 4008, 534 

\bibitem[Fran{\c c}ois et 
al.(2004)]{francois04} Fran{\c c}ois, P., Matteucci, F., Cayrel, R., Spite, M., Spite, F., \& Chiappini, C.\ 2004, \aap, 421, 613 

\bibitem[Freytag et al.(2002)]{frey02} Freytag, B., Steffen, 
M., \& Dorch, B.\ 2002, Astronomische Nachrichten, 323, 213 

\bibitem[Gonz{\'a}lez Hern{\'a}ndez et 
al.(2008)]{jonay08} Gonz{\'a}lez Hern{\'a}ndez, J.~I., et al.\ 2008, \aap, 480, 233 

\bibitem[Gratton et al.(2004)]{gratton04} Gratton, R., Sneden, C., \& Carretta, E.\ 2004, \araa, 42, 385 

\bibitem[Gratton et 
al.(2001)]{gratton01} Gratton, R.~G., et al.\ 2001, \aap, 369, 87 

\bibitem[Israelian 
\& Rebolo(2001)]{israelian01} Israelian, G., \& Rebolo, R.\ 2001, \apjl, 557, L43

\bibitem[Gratton et 
al.(1999)]{gratton99} Gratton, R.~G., Carretta, E., Eriksson, K., \& Gustafsson, B.\ 1999, \aap, 350, 955 

\bibitem[Hubrig et al.(2009)]{hubrig09} Hubrig, S., Castelli, 
F., De Silva, G., Gonzalez, J.~F., Momany, Y., Netopil, M., 
\& Moehler, S.\ 2009, arXiv:0903.5182

\bibitem[Koch 
\& McWilliam(2008)]{koch08} Koch, A., \& McWilliam, A.\ 2008, \aj, 135, 1551 

\bibitem[Korn et al.(2007)]{korn07} Korn, A.~J., Grundahl, F., 
Richard, O., Mashonkina, L., Barklem, P.~S., Collet, R., Gustafsson, B., 
\& Piskunov, N.\ 2007, \apj, 671, 402 

\bibitem[Korn 
\& Ryde(2005)]{korn05} Korn, A.~J., \& Ryde, N.\ 2005, \aap, 443, 1029

\bibitem[{{Kurucz}(1993{\natexlab{b}})}]{kurucz}
{Kurucz}, R. 1993{\natexlab{b}}, SYNTHE Spectrum Synthesis Programs and Line
Data.~Kurucz CD-ROM No.~18.~Cambridge, Mass.: Smithsonian Astrophysical
Observatory, 1993., 18

\bibitem[{{Kurucz}(2005{\natexlab{a}})}]{k05}
{Kurucz}, R.~L. 2005{\natexlab{a}}, Memorie della Societ\`a Astronomica
  Italiana Supplementi, 8, 14
  
  \bibitem[Lanfranchi 
\& Matteucci(2007)]{lanfranchi07} Lanfranchi, G.~A., \& Matteucci, F.\ 2007, \aap, 468, 927
  
 \bibitem[McWilliam(1997)]{mcwilliam97} McWilliam, A.\ 1997, \araa, 35, 503

\bibitem[Meynet \& Maeder(2002)]{MM} Meynet, G., \& Maeder, A.\ 2002, \aap, 390, 561 

\bibitem[Michaud et al.(2008)]{michaud08} Michaud, G., Richer, 
J., \& Richard, O.\ 2008, \apj, 675, 1223 

\bibitem[Nissen et 
al.(2007)]{nissen07} Nissen, P.~E., Akerman, C., Asplund, M., Fabbian, D., Kerber, F., Kaufl, H.~U., \& Pettini, M.\ 2007, \aap, 469, 319

\bibitem[Nissen et 
al.(2004)]{nissen04} Nissen, P.~E., Chen, Y.~Q., Asplund, M., \& Pettini, M.\ 2004, \aap, 415, 993 

\bibitem[Pasquini et 
al.(2005)]{pasquini05} Pasquini, L., Bonifacio, P., Molaro, P., Francois, P., Spite, F., Gratton, R.~G., Carretta, E., \& Wolff, B.\ 2005, \aap, 441, 549 

\bibitem[Ryde(2006)]{ryde06} Ryde, N.\ 2006, \aap, 455, L13 

\bibitem[Ryde 
\& Lambert(2004)]{ryde04} Ryde, N., \& Lambert, D.~L.\ 2004, \aap, 415, 559

\bibitem[Sbordone et 
al.(2007)]{sbordone07} Sbordone, L., Bonifacio, P., Buonanno, R., Marconi, G., Monaco, L., \& Zaggia, S.\ 2007, \aap, 465

\bibitem[{{Sbordone}(2005)}]{sbordone}
{Sbordone}, L. 2005, Memorie della Societ\`a Astronomica Italiana Supplementi, 8,
  61
  
  \bibitem[Sbordone et al.(2005)]{sbordter7} 
Sbordone, L., Bonifacio, P., Marconi, G., Buonanno, R., \& Zaggia, S.\ 2005, \aap, 437, 905 

\bibitem[{{Sbordone} {et~al.}(2004){Sbordone}, {Bonifacio}, {Castelli}, \&
  {Kurucz}}]{s04}
{Sbordone}, L., {Bonifacio}, P., {Castelli}, F., \& {Kurucz}, R.~L. 2004,
  Memorie della Societ\`a Astronomica Italiana Supplementi, 5, 93

\bibitem[Takada-Hidai et al.(2002)]{takada02} Takada-Hidai, M., 
et al.\ 2002, \apj, 573, 614

\bibitem[Takeda et al.(2005)]{takeda} Takeda, Y., Hashimoto, 
O., Taguchi, H., Yoshioka, K., Takada-Hidai, M., Saito, Y., 
\& Honda, S.\ 2005, \pasj, 57, 751 

\bibitem[Tautvai{\v s}ien{\.e} et al.(2004)]{tarantella04} 
Tautvai{\v s}ien{\.e}, G., Wallerstein, G., Geisler, D., Gonzalez, G., 
\& Charbonnel, C.\ 2004, \aj, 127, 373

\bibitem[Ventura et al. (2002)]{ventura02} 
Ventura, P., D'Antona, F., Mazzitelli, I. 2002, A\&A,
393, 215

\bibitem[Wang 
\& Wheeler(2008)]{wang08} Wang, L., \& Wheeler, J.~C.\ 2008, \araa, 46, 433 

\bibitem[Wedemeyer et 
al.(2004)]{wede04} Wedemeyer, S., Freytag, B., Steffen, M., Ludwig, H.-G., \& Holweger, H.\ 2004, \aap, 414, 1121

\bibitem[Wiese et al.(1969)]{wiese} Wiese, W.~L., Smith, 
M.~W., \& Miles, B.~M.\ 1969, NSRDS-NBS, Washington, D.C.: US Department of 
 Commerce, National Bureau of  Standards, |c 1969
 
 \bibitem[Woosley 
\& Heger(2007)]{woosley07} Woosley, S.~E., \& Heger, A.\ 2007, \physrep, 442, 269 

\bibitem[Woosley 
\& Janka(2005)]{woosley05} Woosley, S., \& Janka, T.\ 2005, Nature Physics, 1, 147 

\bibitem[W{\"u}nsch et al.(2008)]{wuensch08} W{\"u}nsch, R., 
Tenorio-Tagle, G., Palou{\v s}, J., \& Silich, S.\ 2008, \apj, 683, 683 

\end{thebibliography}
{}

%

\end{document}